\documentclass[prl,twocolumn,showpacs,superscriptaddress]{revtex4}

\usepackage{graphicx}
\usepackage{bm}
\usepackage[colorlinks,linkcolor=blue,anchorcolor=blue,citecolor=blue]{hyperref}
\usepackage{amssymb}
\usepackage{amsmath}
\usepackage{mathrsfs}
\usepackage[british]{babel}
\usepackage{polski}

\begin{document}

\title{Memory-multi-fractional Brownian motion with continuous correlations}

\author{Wei Wang}
\affiliation{Institute of Physics \& Astronomy, University of Potsdam,
14476 Potsdam, Germany}
\author{Micha{\l} Balcerek}
\affiliation{Faculty of Pure and Applied Mathematics, Hugo Steinhaus Center,
Wroc{\l}aw University of Science and  Technology, Wroc{\l}aw, Poland}
\author{Krzysztof Burnecki}
\affiliation{Faculty of Pure and Applied Mathematics, Hugo Steinhaus Center,
Wroc{\l}aw University of Science and  Technology, Wroc{\l}aw, Poland}
\author{Aleksei V. Chechkin}
\affiliation{Institute of Physics \& Astronomy, University of Potsdam,
14476 Potsdam, Germany}
\affiliation{Faculty of Pure and Applied Mathematics, Hugo Steinhaus Center,
Wroc{\l}aw University of Science and  Technology, Wroc{\l}aw, Poland}
\affiliation{Akhiezer Institute for Theoretical Physics, National Science
Center "Kharkov Institute of Physics and Technology", Kharkov 61108,
Ukraine}
\author{Skirmantas Janu\v{s}onis}
\affiliation{Department of Psychological and Brain Sciences, University
of California, Santa Barbara, Santa Barbara, CA 93106, USA}
\author{Jakub {\'S}l\k{e}zak}
\affiliation{Faculty of Pure and Applied Mathematics, Hugo Steinhaus Center,
Wroc{\l}aw University of Science and  Technology, Wroc{\l}aw, Poland}
\author{Thomas Vojta}
\affiliation{Department of Physics, Missouri University of Science and
Technology, Rolla, MO 65409, USA}
\author{Agnieszka Wy{\l}oma{\'n}ska}
\affiliation{Faculty of Pure and Applied Mathematics, Hugo Steinhaus Center,
Wroc{\l}aw University of Science and  Technology, Wroc{\l}aw, Poland}
\author{Ralf Metzler}
\affiliation{Institute of Physics \& Astronomy, University of Potsdam,
14476 Potsdam, Germany}
\affiliation{Asia Pacific Center for Theoretical Physics, Pohang 37673,
Republic of Korea}

\date{\today}

\begin{abstract}
We propose a generalization of the widely used fractional Brownian motion
(FBM), memory-multi-FBM (MMFBM), to describe viscoelastic or persistent
anomalous diffusion with time-dependent memory exponent $\alpha(t)$
in a changing environment. In MMFBM the built-in, long-range memory is
continuously modulated by $\alpha(t)$. We derive the essential statistical
properties of MMFBM such as response function, mean-squared displacement
(MSD), autocovariance function, and Gaussian distribution. In contrast to
existing forms of FBM with time-varying memory exponents but reset memory
structure, the instantaneous dynamic of MMFBM is influenced by the process
history, e.g., we show that after a step-like change of $\alpha(t)$ the scaling
exponent of the MSD after the $\alpha$-step may be determined by the value
of $\alpha(t)$ before the change. MMFBM is a versatile and useful process
for correlated physical systems with non-equilibrium initial conditions in
a changing environment.
\end{abstract}


\maketitle

The stochastic motion of individual colloidal particles or labeled single
molecules is routinely recorded by single-particle tracking \cite{carlo2}
in soft- and bio-matter systems \cite{pt,saxton,franosch}, i.a., crowded
liquids \cite{matthias,lene1}, cytoplasm of biological cells \cite{yossi,lampo,
weber,yuval,lene}, actively driven tracers \cite{elbaum,seisenhuber,granick},
lipid membranes \cite{diego,natcomm,carlo1}, and porous media \cite{dan}. In
silico, lipid and protein motion \cite{membraneprl,jae_prx,eiji} or internal
protein dynamics \cite{jeremy,eiji} are sampled. On larger scales, motile
cells or small organisms \cite{hapca,gold,carsten}, and animals, e.g., marine
predators or birds \cite{sims,simsalba,ranscience,ranprx,ohad} are traced.
Often the observed motion deviates from Brownian motion with its linear
mean-squared displacement (MSD) $\langle x^2(t)\rangle\simeq t$ and Gaussian
dis\-placement probability density function (PDF) \cite{vankampen}. Instead,
anomalous diffusion with MSD $\langle x^2(t)\rangle\simeq t^{\alpha}$ emerges
\cite{saxton,pt,franosch}, with sub- ($0<\alpha<1$) and superdiffusion
($\alpha>1$) \cite{bouchaud,pt}. Depending on the system, anomalous diffusion
is described by different generalized stochastic models
\cite{bouchaud,igor_sm,pccp,carlo,henrik}.

Two such processes have turned out to be particularly suited to model anomalous
diffusion in a wide range of systems. One is the continuous time random walk, in
which (waiting) times $\tau$ between two successive jumps are randomly distributed
\cite{bouchaud,igor_sm,pccp}. When the PDF of $\tau$ has the scale-free form
$\psi(\tau)\simeq\tau^{-1-\alpha}$ with $0<\alpha<1$, the resulting motion is
subdiffusive \cite{bouchaud,igor_sm,pccp}. Power-law forms for $\psi(\tau)$
were, i.a., measured for colloids in actin gels \cite{weitz,yael}, membrane
channels \cite{diego}, doxorubicin molecules in silica slits \cite{amanda},
ribonucleoproteins in neurons \cite{jaehyong}, foraging birds \cite{ohad},
or in weakly chaotic systems \cite{geisel,swinney}.

The second common anomalous diffusion process is fractional Brownian motion
(FBM) \cite{kolm40,mandelbrot} based on the stochastic equation $dX(t)/dt=
\xi(t)$ driven by fractional Gaussian noise (FGN) with stationary
autocovariance function (ACVF) $\langle\xi(t)\xi(t+\tau)\rangle\sim\frac{1}{2}
\alpha(\alpha-1)K_{\alpha}\tau^{\alpha-2}$ ($0<\alpha\le2$) \cite{weihua,qian}.
Then, $\langle X^2(t)\rangle\simeq K_{\alpha}t^{\alpha}$ with the generalized
diffusivity $K_{\alpha}$ of dimension $\mathrm{length}^2/\mathrm{time}^{\alpha}$.
The ACVF is negative (``antipersistent'') for subdiffusion
and positive (``persistent'') for superdiffusion. Displacement ACVFs consistent
with sub- and superdiffusive FBM were identified, i.a., for tracers in crowded
liquids \cite{weber,lampo,matthias,matthias1,lene1,lene,fox,christine,glebprx},
doxorubicin \cite{amanda}, lipids \cite{membraneprl}, amoeba motion
\cite{christine,glebprx}, and cruising birds \cite{ohad}.
Specifically, subdiffusive FBM models diffusion in viscoelastic systems (cellular
cytoplasm, crowded liquids) \cite{matthias,lene1,weber,lampo,lene,membraneprl},
due to hydrodynamic backflow \cite{alder,mcd,paul,franosch1}, or
``roughness'' in finance \cite{euch,finance}. FBM is intrinsically Gaussian
\cite{kolm40,mandelbrot,qian}, yet, in several viscoelastic systems non-Gaussian
displacement PDFs were found \cite{lampo,natcomm,jae_prx,matthias1,fox}. This
phenomenon (similar to Brownian yet non-Gaussian diffusion \cite{bng,denis})
was ascribed to the systems' heterogeneity and modeled by
superstatistical viscoelastic motion \cite{jakub}, FBM switching
between two diffusivities \cite{matthias1} or featuring a stochastic
(``diffusing'' \cite{diffdiff,gary}) diffusivity \cite{wei,wei1}, and
subordinated FBM \cite{fox}. Random anomalous memory exponents
$\alpha$ were studied in particle ensembles \cite{chechkin,korabel}.

Here we address systems in which the properties of long-range correlated
motions do not vary stochastically but the memory exponent $\alpha$ changes
deterministically over time, $\alpha(t)$.
Examples include smoothly changing viscoelastic environments, e.g., during
biological cell cycles \cite{cycle}, or when pressure and/or concentrations
are changed in viscoelastic solutions \cite{barlow,clemens}. $\alpha(t)$ may
switch more abruptly when the test particle moves across boundaries to
a different environment. Jump-like changes of $\alpha$ may be effected by
binding to larger objects or surfaces \cite{etoc,matthias1} or multimerization
\cite{etoc,heller} of the tracer. Drops in $\alpha$ from superdiffusion with
$\alpha\approx1.8$ to strong subdiffusion $\alpha\approx0.2$ of intracellular
particles were effected by blebbistatin treatment knocking out active
molecular motor action in amoeba cells; after some time, the positive
correlations and thus superdiffusion were restored \cite{christine}. Cellular
sub-micron or micron-sized ``cargo'' transported by molecular motors may
switch between motor-driven transport and rest phases, effecting
repeated sub/superdiffusive switches \cite{motors,igor}. Finally,
crossovers between sub/superdiffusive modes as well as changes in exponents
within sub- or superdiffusion may occur for (intermittent) search of birds
or other animals. We model such situations by a specified protocol $\alpha(t)$
for the memory exponent in our memory-multi-FBM (MMFBM) model, in which the
memory of MMFBM is continuously modulated by $\alpha(t)$. Due to the
uninterrupted memory, the instantaneous dynamic of MMFBM is influenced by the
full history of the process. We study these memory effects on trajectories,
response function, MSD, and ACVF. We show that MMFBM is Gaussian and discuss
relations to other generalized FBM models.

To motivate our approach, consider the simple case of a Brownian particle
with diffusivity $K_1$, released at time $t=0$. At time $t=\tau$ it switches
to a new diffusivity $K_2$, e.g., by crossing to a different environment,
multimerization \cite{heller}, or conformational changes \cite{eiji}. The MSD
of this particle has the form $\langle x^2(t)\rangle=2K_1t$ for $t\le\tau$ and
$=2K_2(t-\tau)+2K_1\tau$ for $t>\tau$. A convenient way to formulate such types
of processes is based on the Wiener process $B(t)$ \cite{vankampen} using
$\tilde{B}(t)=\int_0^t\sqrt {K(s)}dB(s)$. In this formulation $K(s)$
continuously modulates the Wiener increments $dB(s)$ and, e.g., leads
to above MSD.

In a similar fashion we incorporate a time-dependent memory exponent $\alpha(t)$
in FBM. For a physical process initiated at $t=0$ we use L{\'e}vy's
formulation \cite{levy} of non-equilibrated FBM in terms of a (Holmgren)
Riemann-Liouville fractional integral (RL-FBM) \cite{mandelbrot,REM},
\begin{equation}
\label{mmfbm}
X(t)=\int_0^t\sqrt{\alpha(s)}(t-s)^{(\alpha(s)-1)/2}dB(s).
\end{equation}
In standard RL-FBM, the power-law memory kernel with constant exponent $\alpha$
modulates the Wiener increments $dB(s)$ along the path and at long times is
equivalent to integrated FGN. Thus, at any point the process $X(t)$ depends on
its full history. For $\alpha=1$ the kernel vanishes and $X(t)$ is Brownian
motion \cite{mandelbrot}. For changing environments MMFBM incorporates these
changes locally into the memory function, i.e., by variation of how the
correlations of the Wiener increments $dB(s)$ are modulated by $\alpha(s)$
along the path. Thus the uninterrupted history of $\alpha(t)$ is contained yet
the strength of the memory varies throughout the process history. We note that
due to the explicit time dependence of $\alpha(t)$ the noise ACVF is by
construction not stationary. We also note that the structure
(\ref{mmfbm}) for $X(t)$ is similar to time-fractional dynamics of CTRWs with
scale-free waiting time PDF \cite{pccp} and extensions to variable-order with
time-dependent memory exponent \cite{cheso}. We show that MMFBM with its
statistical observables is a meaningful generalization of FBM.

\emph{Response function.} We consider MMFBM (\ref{mmfbm}), that is originally
Brownian (i.e., $\alpha=1$) up to time $\tau$ and then experiences a short period
$\delta$ with exponent $\alpha\neq1$. After $t=\tau+\delta$, the process is
again Brownian. With the increments $X^{\delta}(\tau)=X(\tau+\delta)-X(\tau)$
the response function is
\begin{equation}
\label{response}
\langle X^{\delta}(\tau)X^{\delta}(\tau+T)\rangle=\alpha\delta\frac{\alpha-1}{
2T^{1-\alpha}}\mathrm{B}\left(\frac{\delta}{T};\frac{\alpha+1}{2},1-
\alpha\right)
\end{equation}
for $\delta\to0$, at time $T$ after start of the perturbation with $\alpha$
\cite{sm}. $\mathrm{B}$ is the incomplete Beta function. When $T\to\infty$,
$\langle X^{\delta}(\tau)X^{\delta}(\tau+ T)\rangle\sim\alpha[(\alpha-1)/(
\alpha+1)]\delta^{\alpha/2+3/2}T^{\alpha/2-3/2}$. Thus, even after a long
period $T$ a short perturbation still influences the process, and the sign
of (\ref{response}) depends on whether $\alpha\gtrless1$. For $\alpha=1$
(\ref{response}) is zero, as expected. An example for the scaling
behavior of the response function is shown in Fig.~S7 in SM \cite{sm}.

\begin{figure*}
(a)\includegraphics[width=7.0cm]{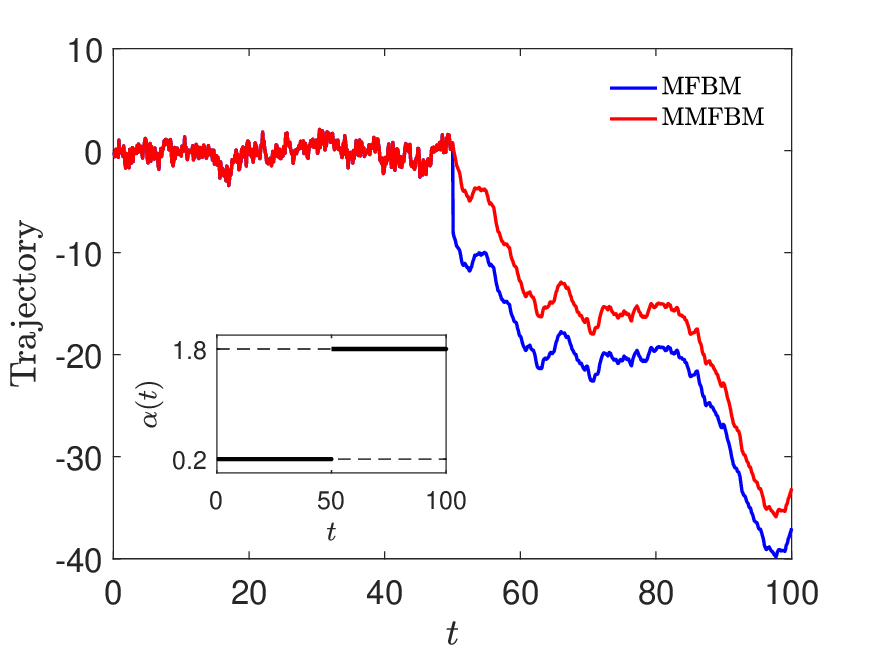}
\hspace*{0.8cm}
(b)\includegraphics[width=7.0cm]{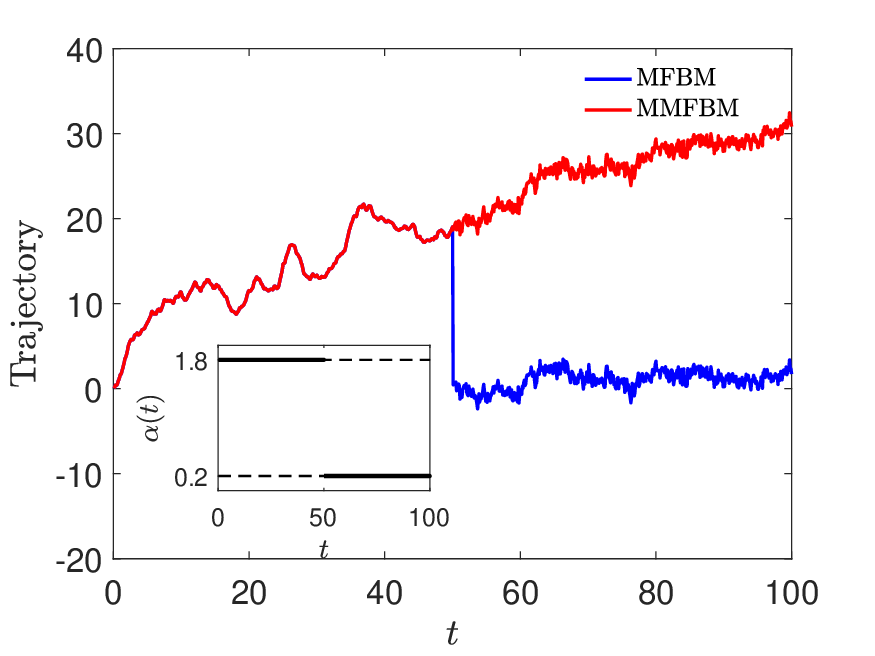}
\caption{Sample trajectories for MMFBM \eqref{mmfbm} (red) and MFBM (S6)
(blue) for step-like protocol (\ref{alphat}) for $\alpha(t)$ with switching
time $\tau=50$ and (a): $\alpha_1=0.2$, $\alpha_2=1.8$; (b): $\alpha_1=1.8$,
$\alpha_2=0.2$. In each panel both trajectories are based on the same
realization of the parental Wiener process. For smooth protocol $\alpha(t)$
see Fig.~S1. On a log-log scale the behavior for smooth and step-like protocol
generally appear quite similar. Note the disparate behavior for $t>\tau$ in
panel (b), see discussion below.}
\label{fig1}
\end{figure*}

In fact, MMFBM (\ref{mmfbm}) is formally similar to definitions in continuous
and discrete time of multifractional FBM (MFBM) \cite{ayache,philippe,surgailis,
ayache1}, a diverse family of processes based on a deterministic $\alpha(t)$
\cite{vehel,stoev}. MFBM and dedicated testing algorithms \cite{michal,agnes}
is used to describe data traffic dynamics \cite{li,riedi}, financial time series
\cite{bianchi}, turbulent dynamics \cite{lee}, or consumer index dynamics
\cite{box}. In most MFBM formulations, it is of interest to describe the
roughness of trajectories and have a globally changing scaling exponent of the
MSD. This is achieved by replacing $\alpha(s)$ in (\ref{mmfbm}) by $\alpha(t)$,
i.e., the Wiener increments $dB(s)$ at time $t$ are modulated by the same
exponent throughout the ``history''. When $\alpha(t)$ changes, the memory of
the correlations is reset and globally replaced by a new weight \cite{REM1}.
The changes of $\alpha(t)$ in MFBM directly affect the MSD, which scales as
$\langle x^2(t)\rangle\simeq t^{\alpha(t)}$. This can be directly seen when
calculating the response function: for MFBM (\ref{response}) is identically
zero, i.e., the reset of the history in MBFM kills any influence of the
perturbation even at short periods $T$. We discuss further differences between
MMFBM and MFBM below, arguing that MMFBM reflects memory properties expected
for long-range correlated dynamics with uninterrupted memory.

\emph{Step-wise $\alpha(t)$-protocol.} To simplify the discussion of the general
properties of MMFBM, we consider a stepwise protocol between two values of
$\alpha$ switching at $t=\tau$,
\begin{eqnarray}
\label{alphat}
\alpha(t)=\left\{\begin{array}{ll}\alpha_1, & t\leq\tau\\
\alpha_2,&t>\tau\end{array}\right.,
\end{eqnarray}
in an unbounded space. More complicated behaviors can be constructed as a
sequence of values $\alpha_i$. Smooth versions of the step-like protocol
(\ref{alphat}) can, e.g., be realized by sigmoid functions (S19) \cite{sm}.
Such forms, however, require numerical analysis. Fig.~\ref{fig1} shows
trajectories of MMFBM for the step-like form (\ref{alphat}), while Fig.~S1
depicts the case of a smooth protocol \cite{sm}. In both Figures we also
show the corresponding MFBM trajectories, for the same parental Wiener
processes $B(s)$. For both processes the roughness change in the
trajectories at $t=\tau$ is distinct. In both cases MMFBM appears more
``continuous''.

\emph{MSD.} With definition (\ref{mmfbm}), the MMFBM-MSD reads
\begin{eqnarray}
\label{MSD-RLMFBM}
\left<X^2(t)\right>=\int_0^t\alpha(s)(t-s)^{\alpha(s)-1}ds,
\end{eqnarray}
due to the independence of the Wiener process at different times. Indeed, the
instantaneous value of the MSD depends on the local modulation by $\alpha(s)$
along the process history. For the stepwise protocol (\ref{alphat}), the MSD
reads
\begin{eqnarray}
\label{msd}
\left<X^2(t)\right>=\left\{\begin{array}{ll}t^{\alpha_1},&t\leq\tau\\
t^{\alpha_1}-(t-\tau)^{\alpha_1}+(t-\tau)^{\alpha_2},&t>\tau\end{array}\right..
\end{eqnarray}
This form contrasts the MFBM result, for which $\left<X^2(t)\right>\propto t^{
\alpha(t)}$ for all $t$, i.e., for step-change (\ref{alphat}) of $\alpha$ the
MSD scaling exponent changes abruptly from $\alpha_1$ to $\alpha_2$ at $t=\tau$:
by memory reset, at time $t$ the history of the previous memory exponents at $s<t$ is
erased in MFBM \cite{vehel,stoev,sm}. We note that in MMFBM even for stepwise
$\alpha(t)$ considered here, the MSD is continuous at $t=\tau$ (the derivative
is continuous for strong memory, $\alpha_1,\alpha_2>1$).

\begin{figure}
\includegraphics[width=7.0cm]{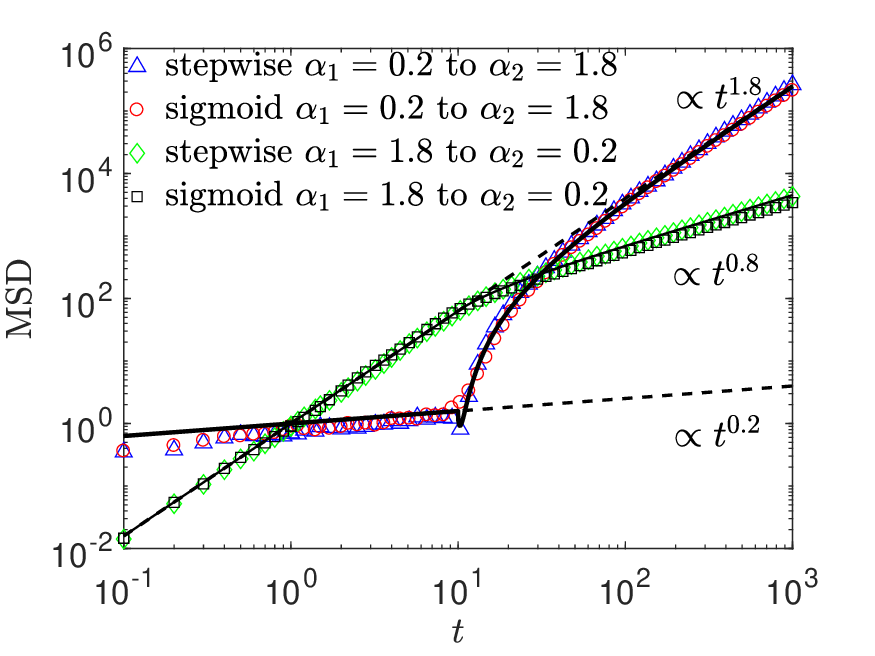}
%(b)\includegraphics[width=7.0cm]{fig2b.eps}
\caption{MSD for MMFBM $X(t)$ \eqref{mmfbm} for step-like and smooth protocol
$\alpha(t)$ for two combinations of $\alpha_1$ and $\alpha_2$ (see legend).
Full lines represent Eq.~(\ref{msd}), symbols represent stochastic simulations.
A comparison with MFBM is shown in Fig.~S2 \cite{sm}.}
\label{fig2}
\end{figure}

The MSD (\ref{msd}) already shows the interesting property that after the
switching point $t=\tau$, both $\alpha_1$ and $\alpha_2$ appear. Expanding
the MSD at long time $t\gg\tau$, we find
\begin{equation}
\label{msdas}
\left<X^2(t)\right>\sim(\alpha_1\tau)t^{\alpha_1-1}+t^{\alpha_2}.
\end{equation}
Fig.~\ref{fig2} shows the time dependence of the MMFBM-MSD for both
step-like and sigmoid protocols, showing perfect agreement with the predicted
asymptotic behavior. In (\ref{msdas}), as long as $\alpha_2>\alpha_1-1$, the
second exponent will eventually dominate the MSD scaling. As shown in SM
\cite{sm}, this convergence can, however, be very slow, much longer than
the switching time $\tau$. Even more, when $\alpha_1>1+\alpha_2$ the MSD
exhibits a continued scaling with $\alpha_1-1$ (as
confirmed in Fig.~\ref{fig2}). In other words, the more superdiffusive
behavior is dominant asymptotically, albeit with the reduced slope $\alpha-1$.
%Qualitatively this result can be understood from the ACVF of fixed-$\alpha$
%FGN, $\langle\xi(t)\xi(t+\Delta)\rangle\simeq\Delta^{\alpha-2}$, that decays
%slower for larger $\alpha$ (and stays constant in the ballistic limit $\alpha
%=2$).

\begin{figure*}
(a)\includegraphics[height=5.6cm]{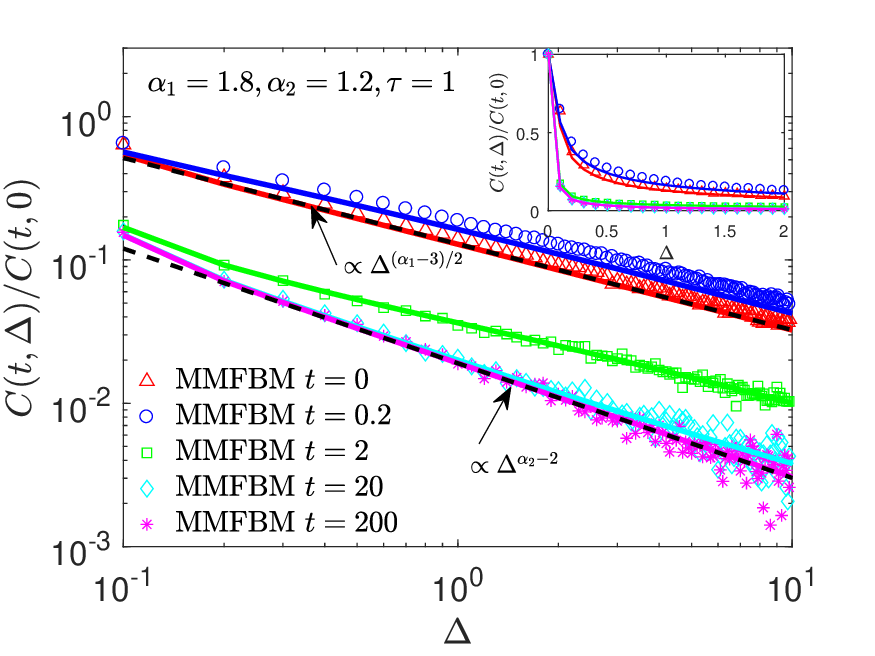}
(b)\includegraphics[height=5.6cm]{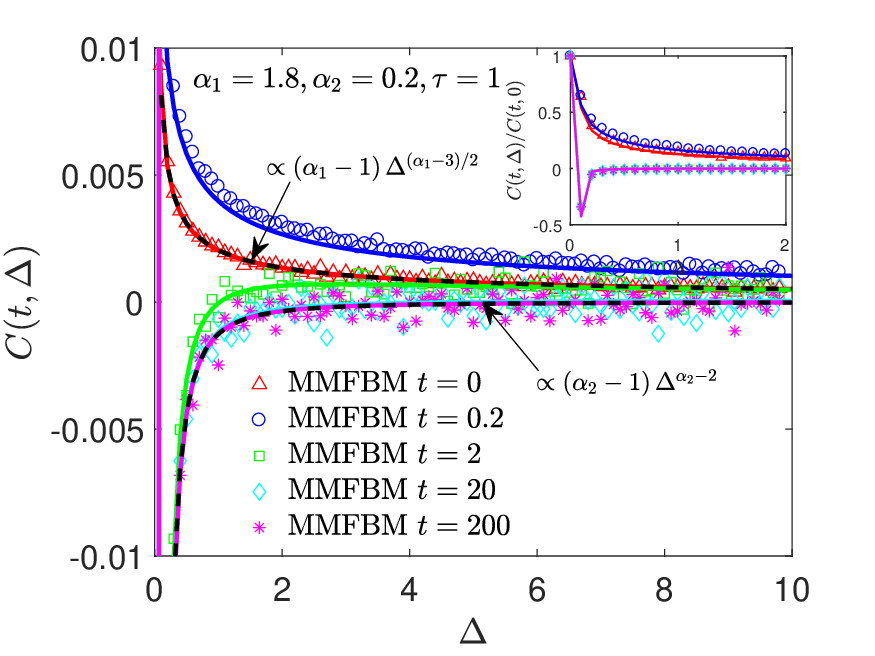}
\caption{Numerical evaluations (lines) and simulations (symbols) for the ACVF
$C(t,\Delta)$ at different times for MMFBM \eqref{mmfbm} with switching time
$\tau=1$ (a): $\alpha_1=1.8$, $\alpha_2=1.2$; (b): $\alpha_1=1.8$, $\alpha_2
=0.2$. The inset in panel (b) shows the numerically obtained form close to
$\Delta=0$, demonstrating the antipersistence at long times $t$, when $\alpha_2
=0.2$ becomes the dominant contribution. Note than in the main panel (b) the
ACVF is shown in non-normalized form for better visibility.}
\label{fig3}
\end{figure*}

\emph{ACVF.} We now study the ACVF, which is defined as
\begin{eqnarray}
\label{acf}
C(t,\Delta)=\langle X^{\delta}(t)X^{\delta}(t+\Delta)\rangle
\end{eqnarray}
with the increments $X^{\delta}(t)=X(t+\delta)-X(t)$. First we consider
short $t$, i.e., the first increment $X^{\delta}(t)$ in (\ref{acf}) is taken
before the
switching time $\tau$ of $\alpha(t)$ in (\ref{alphat}). Obviously, when also
$t+\Delta<\tau$, the ACVF is the same as for RL-FBM with exponent $\alpha_1$
and MFBM (Eq.~(S12) \cite{sm}). This result explicitly depends on both $t$
and $\Delta$, due to the non-stationarity of RL-FBM. When $t=0$ and $\delta
\ll\tau$,
\begin{equation}
\label{acf1}
C(0,\Delta)_{\Delta<\tau}\sim\frac{\alpha_1(\alpha_1-1)\delta^{(\alpha_1+3)
/2}}{\alpha_1+1}\Delta^{(\alpha_1-3)/2}.
\end{equation}
Interestingly, when we correlate increments from before and after the switching
time $\tau$, $t+\Delta>\tau$, the MFBM-ACVF (S11) depends on
both $\alpha_1$ and $\alpha_2$, while for MMFBM the ACVF is exactly that of
\emph{unswitched\/} RL-FBM and solely depends on $\alpha_1$. I.e., for $t=0$ we
recover the form (\ref{acf1}) with $\Delta>\tau$. In fact, this result is not
surprising. MFBM after the switching is fully independent of the process
before the switching, and thus both exponents occur in the ACVF. For MMFBM, in
contrast, the process right after the switching event is still dominated by
the memory from the evolution before the switching. Consequently, the sole
occurrence of $\alpha_1$ is indeed meaningful. At intermediate times, the
MMFBM-ACVF depends on both $\alpha_1$ and $\alpha_2$, as expected (see
(S13)). The result needs to be evaluated numerically.
However, in the limit $t\to\infty$, we expect the ACVF to forget about its
history and solely depend on $\alpha_2$, which is indeed fulfilled,
\begin{equation}
\label{acf2}
C(\infty,\Delta)=\frac{\alpha_2(\alpha_2-1)\Gamma^2\left((\alpha_2+1)/
2\right)\delta^2}{2\Gamma(\alpha_2)\mathrm{sin}(\pi\alpha_2/2)}\Delta^{
\alpha_2-2}.
\end{equation}
Fig.~\ref{fig3} depicts different scenarios for the ACVF (\ref{acf}). Nice
agreement between stochastic simulations and the theoretical results is
observed. Fig.~S3 shows further cases. We also highlight the difference
of the ACVF between the two models when $t$ is close to the switching time
$\tau$ but long lag times $\Delta$ are chosen in SM IV.C \cite{sm}.

\emph{PDF.} The PDF $P_1(x,t)$ of MMFBM for $t\le\tau$ is Gaussian. To compute
the PDF $P_2(x,t)$ after the crossover, we separate the process into two parts,
that are both Gaussian. The PDF of the full process is obtained as
\begin{equation}
P(x,t)=\frac{\exp\left(-x^2/[2(t^{\alpha_1}-(t-\tau)^{\alpha_1}+(t-\tau)^{
\alpha_2})]\right)}{\sqrt{2\pi\left(t^{\alpha_1}-(t-\tau)^{\alpha_1}+(t-
\tau)^{\alpha_2}\right)}},
\end{equation}
which is again a Gaussian process. MMFBM remains Gaussian for any protocol
$\alpha(t)$ of the memory exponent.

\emph{Local regularity.} The self-similarity of a process determines its
fractal (Hausdorff) graph dimension \cite{falconer}. For a Gaussian process
it is determined by the semivariogram (structure function) $\gamma_t(\delta)
=\langle(X^{\delta}(t))^2\rangle$. When $\gamma_t(\delta)\sim D_t\delta^{
\alpha}$, the fractal graph dimension is $2-\alpha/2$. This also
holds for non-stationary increment processes such as RL-FBM \cite{picard}.
For MMFBM with protocol (\ref{alphat}) for $t<\tau$, $X(t)$ is identical to
RL-FBM, so this part of the trajectory has fractal dimension $2-\alpha_1/2$.
After the switch it can be shown that the trajectory has fractal dimension
$2-\alpha_2/2$. For any graph containing a piece before and after $\tau$,
the lower fractional index and thus the higher fractal dimension dominates
\cite{unp}.

\emph{Conclusions.} FBM is a widely used process to describe anomalous
diffusion in soft- and bio-matter systems. It is characterized by long-ranged,
positive or negative correlations in time. Yet many
real-world systems exhibit changes in the anomalous diffusion exponent (and
thus the memory exponent modulating the correlations in the motion) as function of
time. Prime examples include environments, in which particles cross between
areas of different viscoelastic properties or when the degree of crowding
is controlled. Cargo being pulled intermittently by molecular motors switch
between sub- and superdiffusion in cells, and search strategies of birds
with correlated increments may vary over time as they switch their motion
mode in response to the environment, time of day, or season. Tracers in
fluidic setups that modulate between effectively three- and two-dimensional
embedding should change the exponent of the power-law Basset force.
 In finance the
instantaneous degree of roughness of the trading data may vary during the daily
rhythm, following interventions in the market, or due to longer-lasting
events such as pandemics, wars, or vacation times. Long-range correlated
processes such as viscoelastic anomalous diffusion necessarily feature
effects of memory of the entire dynamics in physical observables such as
the MSD or the ACVF, both of which can be measured.

We here introduced MMFBM as a generalization of FBM to a deterministic form
$\alpha(t)$ of the memory exponent. In the correlation integral $\alpha(s)$
locally modulates the Wiener increments $dB(s)$ and thus contributes to the
correlation history of the process. The MSD, and the ACVF of MMFBM exhibit
crossovers carrying explicit information from the process prior to switching.
This contrasts MFBM, which resets the previous history globally, as seen in the
MSD $\langle x^2(t)\rangle\simeq t^{\alpha(t)}$, that solely depends on the
instantaneous value of $\alpha$ at process time $t$. While this reset of
correlation history is irrelevant
when discussing the instantaneous roughness of a trajectory, for a physical
process with long-range correlations this point is crucial
when the correlations are directly probed, e.g., in single particle tracking
experiments. Here, MMFBM appears physically consistent. We hope that
MMFBM will find wide use in soft- and bio-matter systems, finance, ecology,
etc. MMFBM will also extend the arsenal of generalized stochastic processes in
data analysis \cite{carlo,henrik}.

Our discussion was based on non-stationary RL-FBM. MMFBM is thus useful for the
description of typical physical systems initiated at $t=0$ that first have
to equilibrate. We demonstrated that at sufficiently long times asymptotic
stationarity is restored. It will be interesting to see how MMFBM is modified
in the fully stationary limit, i.e., generalizing Mandelbrot-van Ness FBM
for systems, that are equilibrated at
the start of the measurement. We note that apart from using a purely
time-dependent protocol $\alpha(t)$ corresponding to deterministic
modifications of the system, it will be interesting to consider scenarios of
space-varying scaling exponents in a heterogeneous, quenched system, as well
as to combine a protocol $\alpha(t)$ with a time dependence of the (generalized)
diffusion coefficients as observed in \cite{matthias1}. Moreover, non-Gaussian
extensions of MMFBM should be studied, as well as effects of cutoffs or tempering
\cite{daniel} of the correlations. Finally it should be studied how the
non-standard behavior of FBM \cite{vojta,tobias} next to boundaries is modified
for MMFBM, relevant, e.g., for growing serotonergic fibers in inhomogeneous
brain environments \cite{skirmantas}.

\begin{acknowledgments}
We acknowledge support from German Science Foundation (DFG grant ME 1535/12-1)
and NSF-BMBF CRCNS (grant 2112862/STAXS). AVC acknowledges support by the
Polish National Agency for Academic Exchange (NAWA). AW was supported
by the Polish National Center of Science (Opus 2020/37/B/HS4/00120).
KB acknowledges support through Beethoven Grant DFG-NCN 2016/23/G/ST1/04083.
\end{acknowledgments}

\clearpage

\onecolumngrid 

\setcounter{figure}{0}
\renewcommand{\thefigure}{S\arabic{figure}}
\setcounter{equation}{0}
\renewcommand{\theequation}{S\arabic{equation}}

\center{\large\textbf{Supplementary Material:\\[0.2cm]
Memory-multi-fractional Brownian motion with continuous correlations}}

\section{L{\'e}vy's Riemann-Liouville-FBM}

A well-known representation of fractional Brownian motion (FBM) is attributed
by Mandelbrot \cite{mandelbrot} to Paul L{\'e}vy \cite{levy}. It is given by
the Holmgren-Riemann-Liouville fractional integral \cite{mandelbrot}
\begin{eqnarray}
\label{rl-fbm}
B_{\alpha}(t)=\sqrt{\alpha}\int_0^t(t-s)^{(\alpha-1)/2}dB(s)
\end{eqnarray}
$B(t)$ is standard Brownian motion and $\alpha\in(0,2]$. It is easy to show
that the MSD of $B_{\alpha}(t)$ yields as $\langle B_\alpha^2(t)\rangle=t^{
\alpha}$. This exact equality follows from the choice of the prefactor in
definition \eqref{rl-fbm} \cite{chechkin}.

With the increments of RL-FBM for disjoint intervals $[t,t+\delta]$ and $[t+
\Delta,t+\Delta+\delta]$,
\begin{equation}
B^\delta_\alpha(t)=B_\alpha(t+\delta)-B_\alpha(t),\,\,\,B^\delta_\alpha(t+
\Delta)=B_\alpha(t+\Delta+\delta)-B_\alpha(t+\Delta),
\end{equation}
the ACVF of RL-FBM is given by
\begin{equation}
C_B(t,\Delta)=\langle B^\delta_\alpha(t)B^\delta_\alpha(t+\Delta)\rangle.
\end{equation}
For $\Delta\gg\delta$, after some transformations we obtain the ACVF
\begin{eqnarray}
C_B(t,\Delta)&\approx&\frac{\alpha(\alpha-1)(3-\alpha)\delta^2}{4}\int_0^tq^{
(\alpha-1)/2}(q+\Delta)^{(\alpha-5)/2}dq+\frac{\alpha(\alpha-1)\delta}{2}\int_
t^{t+\delta}q^{(\alpha-1)/2}(q+\Delta)^{(\alpha-3)/2}dq
\nonumber\\
&&\hspace*{-1.4cm}=\frac{\alpha(\alpha-1)(3-\alpha)\delta^2}{4}
\mathrm{B}\left(\frac{t/\Delta}{1+t/\Delta};
\frac{\alpha+1}{2},2-\alpha\right)\Delta^{\alpha-2}+\frac{\alpha(\alpha-1)\delta
}{2}\int_t^{t+\delta}q^{(\alpha-1)/2}(q+\Delta)^{(\alpha-3)/2}dq,
\label{acf-rl-fbm}
\end{eqnarray}
where $\mathrm{B}(z;a,b)$ is the incomplete Beta function \cite{abramowitz}
\begin{equation}
\label{incomplete-beta-function}
\mathrm{B}(z;a,b)=\int_0^zs^{a-1}(1-s)^{b-1}ds.
\end{equation}
RL-FBM has non-stationary increments at any given time $t$, i.e.,
it does not solely depend on the time lag $\Delta$.

\section{MFBM}

\begin{figure*}
(a)\includegraphics[width=8cm]{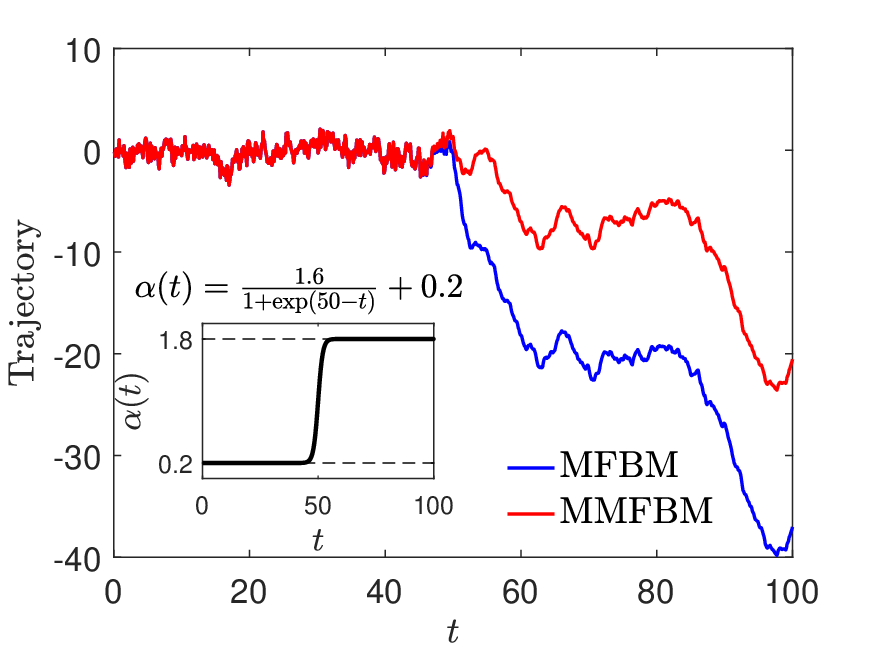}
(b)\includegraphics[width=8cm]{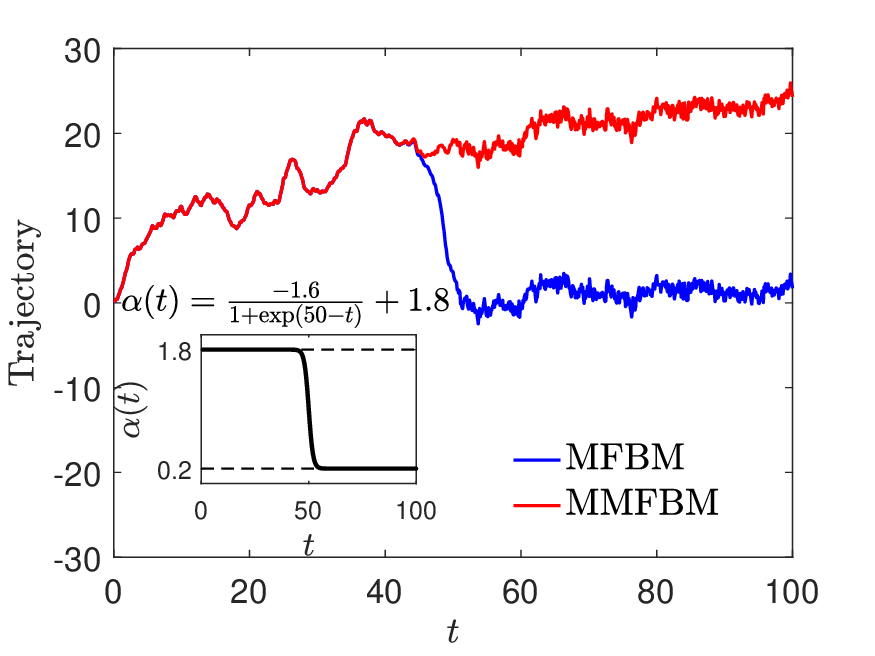}
\caption{Trajectories for MMFBM (1) and MFBM \eqref{rl-mfbm} for a smooth
protocol (\ref{expprot}) for $\alpha(t)$ with switching time $\tau=50$ and (a):
$\alpha_1=0.2$, $\alpha_2=1.8$; (b): $\alpha_1=1.8$, $\alpha_2=0.2$. In each
panel both trajectories are based on the same realization of the underlying
Wiener process. The time series of the parental Wiener increments is also the
same as for Fig.~1, in which we show trajectories produced by a step-like
protocol (2).}
\label{fig:trajs}
\end{figure*}

A direct generalization of FBM to multifractional Brownian motion (MFBM) is
to replace $\alpha$ by an explicitly time-dependent function $\alpha(t)$. In
comparison to mathematical literature (see, e.g., \cite{stoev,vehel,ayache})
we base the generalization on RL-FBM (\ref{rl-fbm}) with the square-root
prefactor,
\begin{equation}
\label{rl-mfbm}
Y(t)=\sqrt{\alpha(t)}\int_0^t(t-s)^{(\alpha(t)-1)/2}dB(s).
\end{equation}
In MFBM, the long-range correlations are reset, as only the
instantaneous value of $\alpha$ at time $t$ is considered in (\ref{rl-mfbm}),
and the MSD scales like
\begin{equation}
\label{msdmfbm}
\langle Y^2(t)\rangle=t^{\alpha(t)}.
\end{equation}
Trajectories for MFBM and MMFBM for the step-like protocol (3) are shown in
Fig.~1, and for a smooth protocol in Fig.~\ref{fig:trajs}. While for the
smooth protocol the discontinuity seen in Fig.~1 is remedied, the general
shape of the trajectories in both cases are quite similar.

\section{MSD of MFBM}

\begin{figure}
(a)\includegraphics[width=8cm]{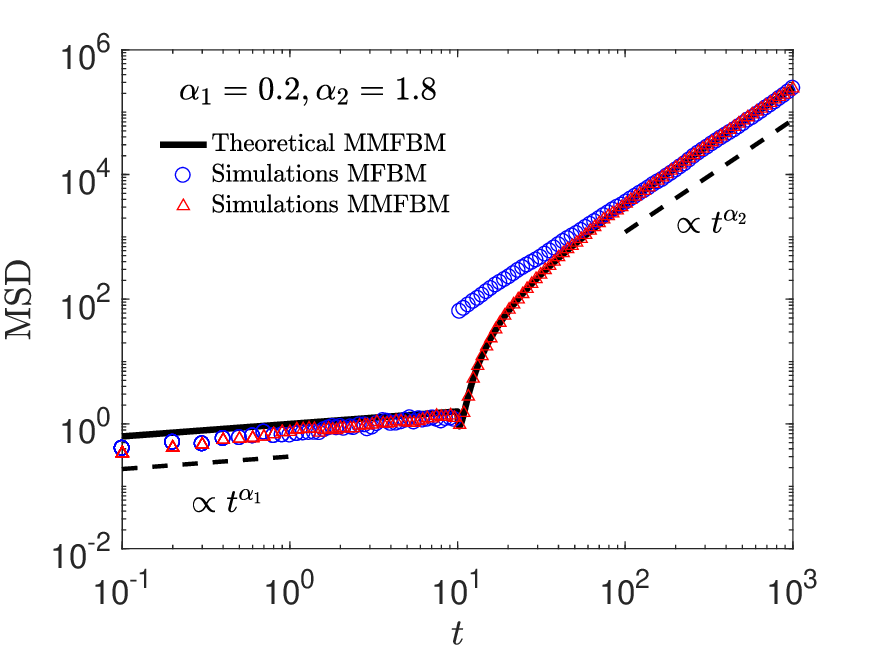}
(b)\includegraphics[width=8cm]{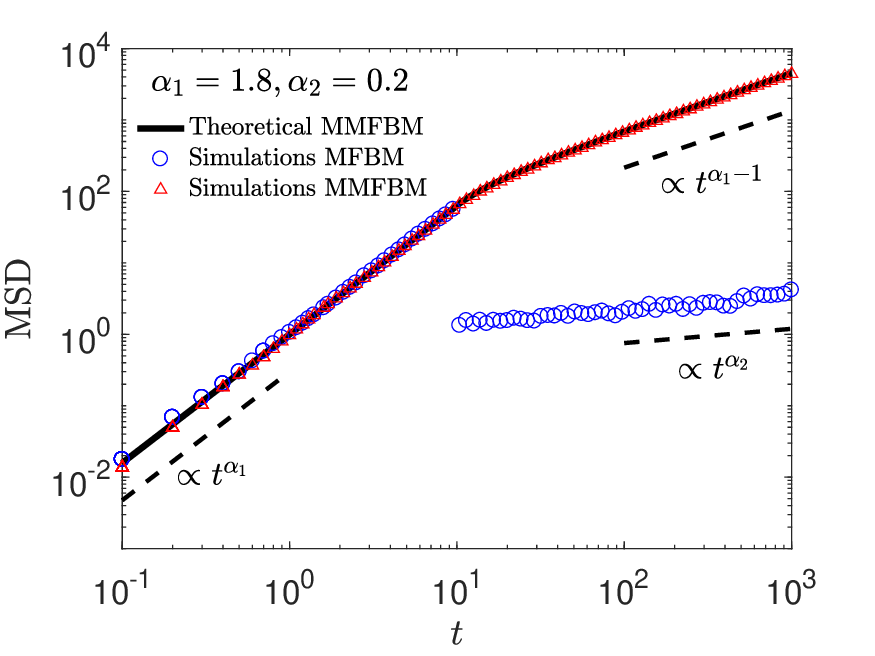}
\caption{MSDs for MFB and MMFBM with switching time $\tau=10$: (a) $\alpha_1=
0.2$, $\alpha_2=1.8$,(b) $\alpha_1=1.8$, $\alpha_2=0.2$.
The theoretical MSD (5) of MMFBM is represented by solid black curves.
Note that the MMFBM-MSD is always continuous, while the derivative is
continuous when $\alpha_1,\alpha_2>1$. For MFBM the MSD is always
discontinuous.}
\label{fig:msd}
\end{figure}

The MSD of MFBM \eqref{rl-mfbm} with step-like anomalous diffusion exponent
jumping from $\alpha_1$ to $\alpha_2$ at $t=\tau$, yields in the form
\begin{equation}
\label{msd-rl-mfbm}
\langle Y^2(t)\rangle=\left\{\begin{array}{ll}t^{\alpha_1},& t\leq\tau\\[0.2cm]
t^{\alpha_2},& t>\tau\end{array}\right.
\end{equation}
Indeed, for $t>\tau$, solely the value $\alpha_2$ appears, due to the
reset correlations of MFBM.
The MSDs for MFBM and MMFBM are displayed in Fig.~\ref{fig:msd} along with
stochastic simulations.

\section{ACVF for MMFBM and MFBM}

We define the increment of MMFBM as $X^\delta(t)=X(t+\delta)-X(t)$. The ACVF
is given by $C_X(t,\Delta)=\langle X^\delta (t)X^\delta(t+\Delta)\rangle$,
where the time step $\delta$ is taken to be small, $\delta\ll\Delta,\tau$.
Similarly, the ACVF for MFBM \eqref{rl-mfbm} is $C_Y(t,\Delta)$. In the limits
of short and long times $t$ analytical results can be obtained for the ACVFs
for step-like protocol of $\alpha$.

\subsection{Short time limit of the ACVF}

We first consider $t<\tau$. Then we distinguish two cases:

(i) When $t+\Delta<\tau$ we have the same ACVF \eqref{acf-rl-fbm} as that of
RL-FBM. In particular, when $t=0$ and $\Delta\gg\delta$, we have the ACVF for
both MMFBM and MFBM according to
\begin{equation}
\label{acf-rl-mfbm-with-memory-shot-time-1}
C_X(0,\Delta)_{\Delta<\tau}=C_Y(0,\Delta)_{\Delta<\tau}=C_B(0,\Delta|\alpha_1)
\sim\frac{\alpha_1(\alpha_1-1)\delta^{(\alpha_1+3)/2}}{\alpha_1+1}\Delta^{(
\alpha_1-3)/2}.
\end{equation}

\begin{figure}
(a)\includegraphics[width=8cm]{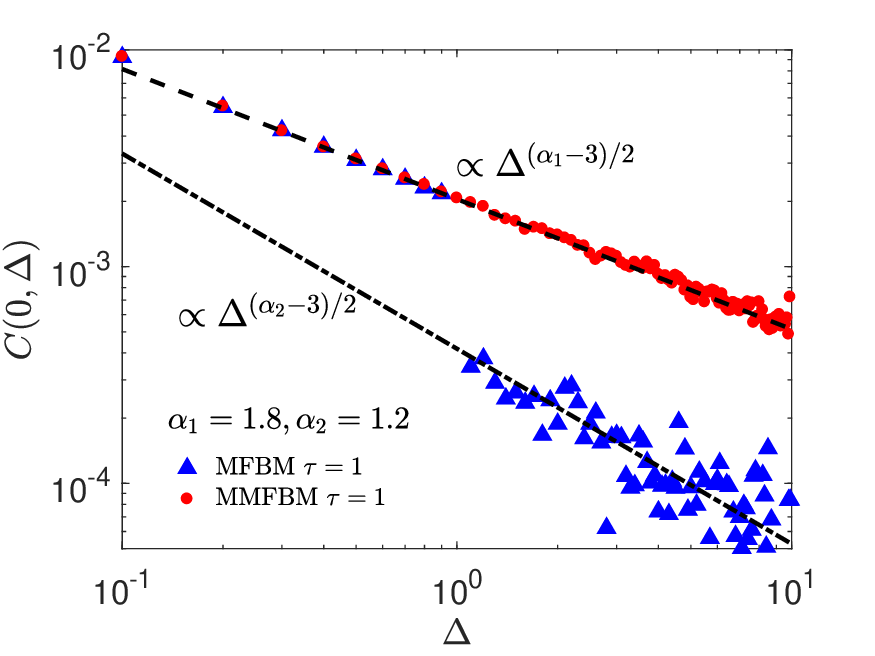}
(b)\includegraphics[width=8cm]{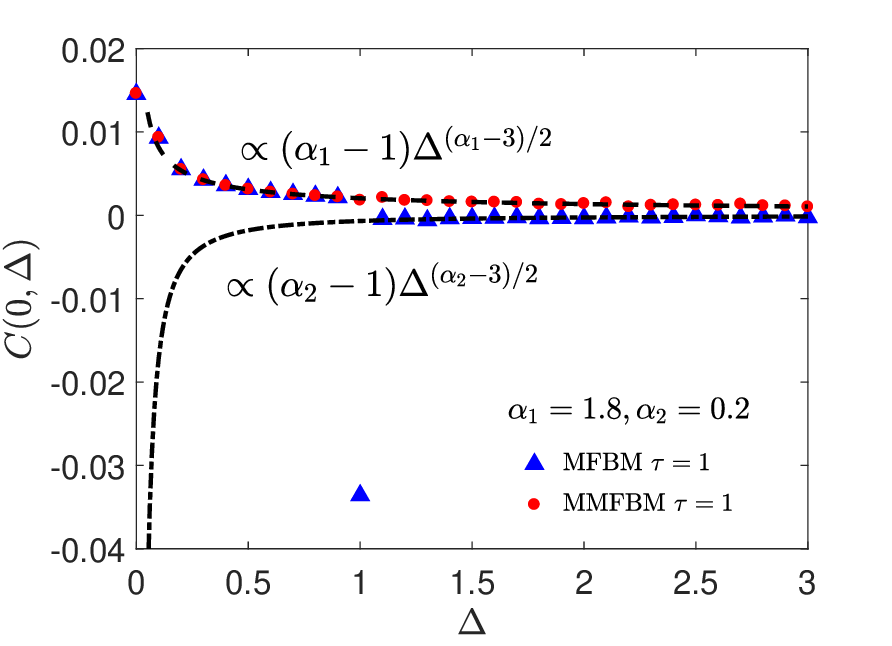}
\caption{ACVF for MFBM and MMFBM with switching time $\tau=1$.
The theoretical ACVF of classic MFBM
\eqref{acf-rl-mfbm-with-memory-shot-time-1} switches to
\eqref{acf-rl-mfbm-short-time-2} when the anomalous diffusion exponent switches.
The ACVF $C(0,\Delta)$ of MMFBM \eqref{acf-rl-mfbm-with-memory-shot-time-1},
\eqref{acf-rl-mfbm-with-memory-shot-time-2} remains the same depending on
$\alpha_1$.} 
\label{fig:acf-1}
\end{figure}

(ii) When $t+\Delta>\tau$, the increment $X^\delta(t+\Delta)$ of MMFBM is
measured after switching and $X^\delta(t)$ before switching.
Multiplying the two increments and averaging over the realizations, the ACVF
is independent of $\alpha_2$ and coincides with the results \eqref{acf-rl-fbm}
of RL-FBM. For $t=0$ and $\Delta\gg\delta$,
\begin{equation}
\label{acf-rl-mfbm-with-memory-shot-time-2}
C_X(0,\Delta)_{\Delta>\tau}\sim\frac{\alpha_1(\alpha_1-1)\delta^{(\alpha_1+3)
/2}}{\alpha_1+1}\Delta^{(\alpha_1-3)/2}.
\end{equation}
In contrast, for MFBM the increment after switching is given by $Y(t+\Delta+
\delta)-Y(t+\Delta)$ and the ACVF for MFBM depends on both $\alpha_2$ and
$\alpha_1$,
\begin{eqnarray}
\nonumber
C_Y(t,\Delta)_{t+\Delta>\tau}&=&
\frac{\sqrt{\alpha_1\alpha_2}(\alpha_2-1)(3-\alpha_2)\delta^2}{4}\mathrm{B}
\left(\frac{t/\Delta}{1+t/\Delta};\frac{\alpha_1+1}{2},2-\frac{\alpha_1+\alpha_2}{2}\right)
\Delta^{(\alpha_1+\alpha_2)/2-2}\\
&&+\frac{\sqrt{\alpha_1\alpha_2}(\alpha_2-1)\delta}{2}\int_t^{t+\delta}q^{(
\alpha_1-1)/2}(q+\Delta)^{(\alpha_2-3)/2}dq.
\label{acf_mfbm}
\end{eqnarray}
When $t=0$ and $\Delta\gg\delta$, we have the MFBM-ACVF
\begin{equation}
\label{acf-rl-mfbm-short-time-2}
C_Y(0,\Delta)_{\Delta>\tau}\sim\frac{\sqrt{\alpha_1\alpha_2}(\alpha_2-1)\delta^{
(\alpha_1+3)/2}}{\alpha_1+1}\Delta^{(\alpha_2-3)/2}.
\end{equation}
The ACVF of MMFBM and MFBM are shown in Fig.~\ref{fig:acf-1}. When the anomalous
diffusion exponent switches from $\alpha_1$ to $\alpha_2$, the ACVF for $t=0$ of
MFBM crosses over from the scaling $\Delta^{(\alpha_1-3)/2}$ to $\Delta^{(\alpha
_2-3)/2}$, while the MMFBM-ACVF retains the scaling $\Delta^{(\alpha_1-3)/2}$.
This clearly shows the uninterrupted memory of MMFBM, in contrast to MFBM.

\subsection{Long time limit of the ACVF}

When $t>\tau$ both increments are observed after the switching of $\alpha$.
After some transformations we obtain the ACVF
\begin{equation}
\label{acvflong}
C_X(t,\Delta)=f\left(\alpha_1,\frac{t}{\Delta}\right)-f\left(\alpha_1,\frac{
t-\tau}{\Delta}\right)+f\left(\alpha_2,\frac{t-\tau}{\Delta}\right)+g\left(
\alpha_1,\frac{t}{\Delta}\right)-g\left(\alpha_1,\frac{t-\tau}{\Delta}\right)
+g\left(\alpha_2,\frac{t-\tau}{\Delta}\right),
\end{equation}
where
\begin{equation}
f(\alpha,s)=\frac{\alpha(\alpha-1)\delta^2}{2}s^{(\alpha-1)/2}(1+s)^{(\alpha-3)
/2}\Delta^{\alpha-2},\,\,\,
g(\alpha,s)=\frac{\alpha(\alpha-1)(3-\alpha)\delta^2}{4}\mathrm{B}\left(\frac{s
}{1+s};
\frac{\alpha+1}{2},2-\alpha\right)\Delta^{\alpha-2}.
\end{equation}
When $t\to\infty$,
\begin{equation}
\label{acf-rl-mfbm-with-memory-long-time}
C_X(\infty,\Delta)=g(\alpha_2,\infty)=\frac{\alpha_2(\alpha_2-1)\Gamma^2
\left((\alpha_2+1)/2\right)\delta^2}{2\Gamma(\alpha_2)\mathrm{sin}(\pi\alpha
_2/2)}\Delta^{\alpha_2-2}.
\end{equation}

For MFBM, the increments depend locally on the Hurst exponents at time $t$.
The ACVF of MFBM is the same as that of FBM with the same $\alpha_2$ at time
$t$ after switching, and when $t\to\infty$,
\begin{equation}
\label{acf-rl-mfbm-long-time}
C_Y(\infty,\Delta)=\frac{\alpha_2(\alpha_2-1)\Gamma^2\left((\alpha_2+1)/
2\right)\delta^2}{2\Gamma(\alpha_2)\mathrm{sin}(\pi\alpha_2/2)}\Delta^{
\alpha_2-2}.
\end{equation}
As it should be, at extremely long times beyond the switching time, the
ACVFs of RL-FBM, MFBM, and MMFBM converge to the same behavior.

\subsection{Long lag time limit of ACVF for times around the switching time}

\begin{figure}
\centering
(a)\includegraphics[width=0.47\linewidth]{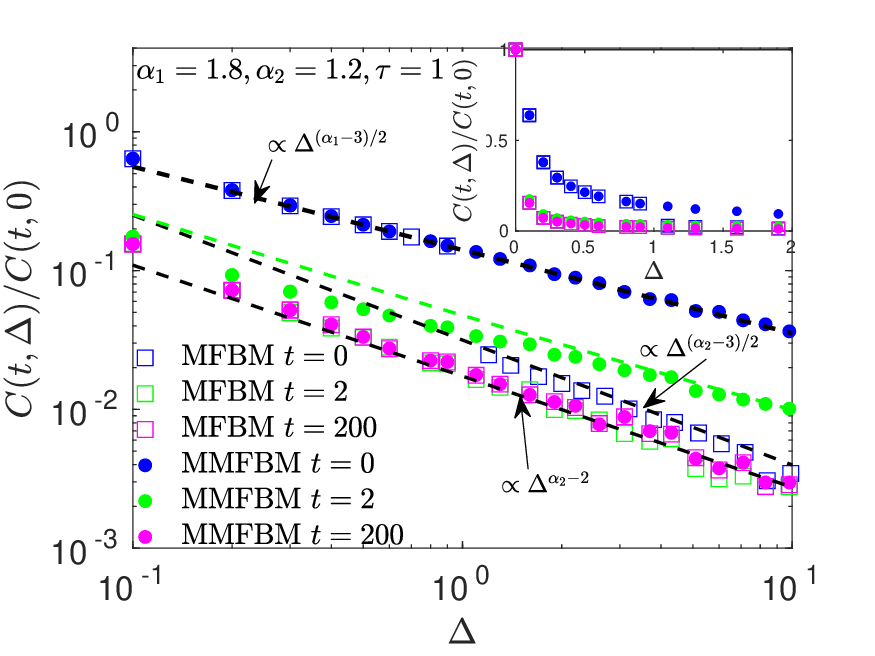}
(b)\includegraphics[width=0.47\linewidth]{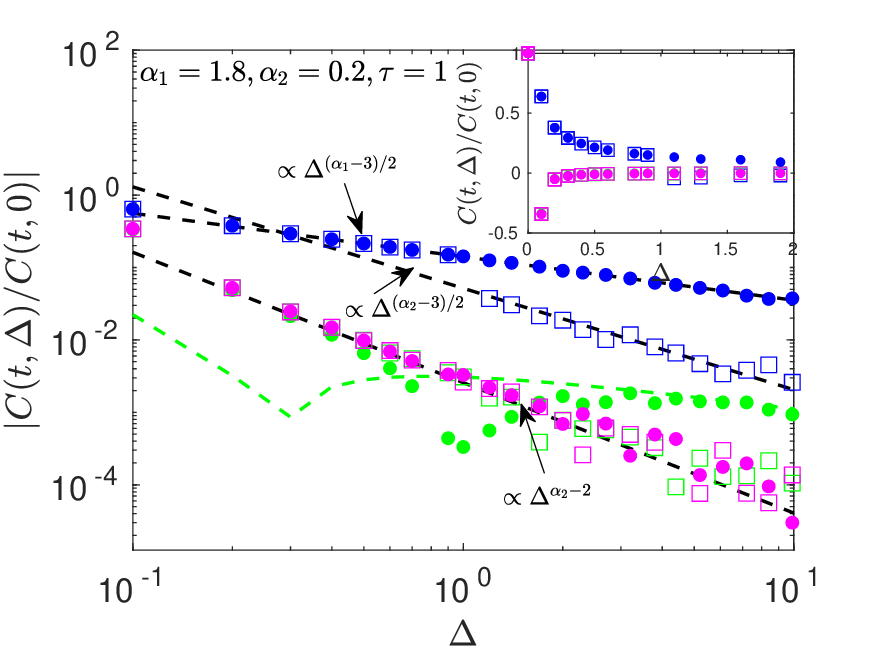}
\caption{Simulations of the ACVFs for MFBM and MMFBM with switching time $\tau
=1$ and (a) $\alpha_1=1.8,\alpha_2=1.2$, (b) $\alpha_1=1.8,\alpha_2=0.2$. The
ACVF of MFBM with $t=0$ is discontinuous at the switching time $\tau$ and
switches to $\Delta^{(\alpha_2-3)/2}$ from $\Delta^{(\alpha_1-3)/2}$ (blue
squares). In contrast, the ACVF of MMFBM is always continuous with scaling
$\Delta^{(\alpha_1-3)/2}$ (blue circles), demonstrating the influence of the
memory from the process before switching. The ACVF of MFBM at around the
switching time $t\approx\tau$ solely depends on instantaneous the exponent
$\alpha_2$ (green squares), Eq.~(\ref{immediate-ACVF-MFBM}), while the ACVF
of MMFBM depends on both $\alpha_1$ and $\alpha_2$ (green circles) with the
long lag time behaviors given by the combination the two scaling behaviors
in Eq.~(\ref{immediate-ACVF-MMFBM}) (green curves). The ACVFs converge at
long time $t$ to the scaling $\Delta^{\alpha_2-2}$.}
\label{newacvf}
\end{figure}

We finally consider the limit of long lag times, $\Delta\gg t$, while the time
$t$ is taken to be close to the switching time, $t\gtrapprox\tau$. As the
incomplete Beta function $\mathrm{B}(s;a,b)\sim s^a/a$ when $s\ll1$, the
function $g(\alpha,s)\simeq s^{(\alpha+1)/2}\Delta^{\alpha-2}$ with $s=t/\Delta$
or $s=(t-\tau)/\Delta$ in Eq.~(\ref{acvflong}) can be neglected in comparison
with $f(\alpha,s)$, and one can approximate the ACVF as
\begin{equation}
C_X(t,\Delta)\sim d_1\Delta^{(\alpha_1-3)/2}+d_2\Delta^{(\alpha_2-3)/2},             \label{immediate-ACVF-MMFBM}
\end{equation}
where $d_1=\frac{1}{2}\alpha_1(\alpha_1-1)\delta^2(t^{(\alpha_1-1)}-(t-\tau)
^{(\alpha_1-1)})$ and $d_2=\frac{1}{2}\alpha_2(\alpha_2-1)\delta^2(t-\tau)^{(
\alpha_2-1)}$. Thus the ACVF is a combination of two scaling behaviors $\Delta
^{(\alpha_1-3)/2}$ and $\Delta^{(\alpha_2-3)/2}$. The former scaling is
inherited from before the switching, i.e., is caused by the memory of MMFBM,
and the latter emerges with the instantaneous exponent after switching. In
Fig.~\ref{newacvf}, the combination of the scaling of the ACVF is show by the
green dashed curves. For the case in the left panel of Fig.~\ref{newacvf}, the
intermediate-scaling behavior predicted by Eq.~(\ref{immediate-ACVF-MMFBM}) is
close to the simulated behavior. In the right panel, the minimum of the ACVF
is not captured well, however, we see convergence at sufficiently long lag times.

In contrast to MMFBM, the ACVF of MFBM solely depends on the instantaneous
exponent $\alpha_2$ and is given by
\begin{eqnarray}
C_Y(t,\Delta)=C_B(t,\Delta|\alpha_2)\sim\frac{\alpha_2\left(\alpha_2-1\right)
\Gamma^2\left(\left(\alpha_2+1\right)/2\right)\delta^2}{2\Gamma\left(\alpha_2
\right)\sin\left(\pi\alpha_2/2\right)}\Delta^{\alpha_2-2},
\label{immediate-ACVF-MFBM}
\end{eqnarray}
where $C_B(t,\Delta|\alpha_2)$ is the ACVF of RL-FBM with exponent $\alpha_2$.

Simulations for both $C_X(t,\Delta)$, Eq.~(\ref{immediate-ACVF-MMFBM}), and
$C_Y(t,\Delta)$, Eq.~(\ref{immediate-ACVF-MFBM}), around switching time are
represented by the green symbols in Fig.~\ref{newacvf}. We note that the
simulations at long lag time $\Delta$ in the right panel exhibit more
pronounced fluctuations due to the subdiffusive behaviors after switching.

\section{Smooth switching of anomalous diffusion exponent}

While the stepwise protocol (2) used here simplifies the analytical
calculation, it is interesting to consider smooth variations. We here briefly
study the exponentially switching anomalous diffusion exponent
\begin{eqnarray}
\label{expprot}
\alpha(t)=\frac{\alpha_2-\alpha_1}{1+\exp\Big(-(t-\tau)/T\Big)}
+\alpha_1,
\end{eqnarray}
where $T$ is some characteristic time measuring how fast the exponent switches
from $\alpha_1$ to $\alpha_2$ around $t=\tau$. At short times $t\ll\tau$, we
see that $\alpha(t)\approx\alpha_1$ while at long times $t\gg\tau$, $\alpha(t)
\approx\alpha_2$. Numerically evaluated trajectories and MSDs for MFBM and MMFBM
with smoothly switching exponent (\ref{expprot}) are displayed
in Figs.~\ref{fig:trajs} and 2.

\section{Physically asymptotic behavior}

As we showed in the main text, a new scaling of the MSD of MMFBM with the
power-law $t^{\alpha_1-1}$ emergies at long times when the memory before
switching is strong, $\alpha_1>\alpha_2+1$. Otherwise, when this inequality
is not fulfilled, the MSD of MMFBM converges to MFBM in the mathematically
asymptotic limit $t\to\infty$. The natural question that arises: what is the
physically measurable time scale, after which the MSDs of the two models
converge? The physical time scale to compare with is given by the switching
time $t=\tau$. To reveal the characteristic time for the convergence of the
two MSDs, we test the ratio of the difference of the MSDs of the two models
(MMFBM Eq.~(1) and MFBM Eq.~(\ref{rl-mfbm})) to that of MFBM at time $t\gg\tau$,
\begin{eqnarray} 
\mathrm{Error}&=&\frac{\left<X^2(t)\right>-\left<Y^2(t)\right>}{\left<Y^2(t)
\right>}=\frac{\left(t^{\alpha_1}-(t-\tau)^{\alpha_1}+(t-\tau)^{\alpha_2}
\right)-t^{\alpha_2}}{t^{\alpha_2}}\nonumber\\
&\sim&\frac{\alpha_1\tau}{t^{1+\alpha_2-\alpha_1}}-\frac{\alpha_2\tau}{t}.      \label{error}
\end{eqnarray}
This allows us to distinguish three cases:

(i) When $\alpha_1>\alpha_2+1$, the relative deviation of the MSD of MMFBM
to MFBM grows with the power $t^{\alpha_1-\alpha_2-1}$. In this case, the
MMFBM never converges to MFBM at long times. Instead, the new scaling $t^{
\alpha_1-1}$ of the MSD emerges, which corresponds to Eq.~(6) and is shown
in Fig.~2 in the main text.

\begin{figure}
\centering
\includegraphics[width=0.47\linewidth]{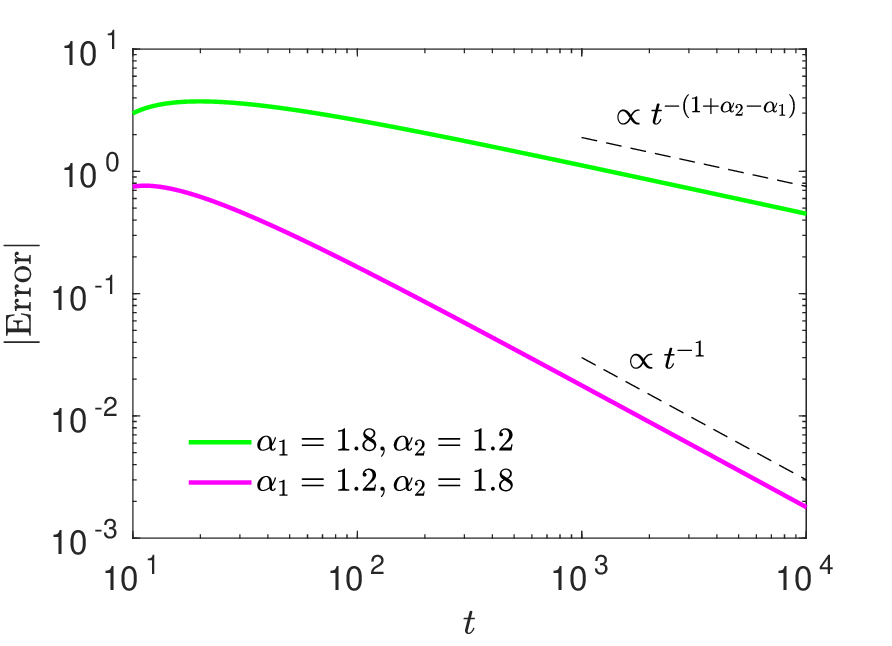}
\caption{Error (\ref{error}) of the MSD from MMFBM to MFBM. While
for $\alpha_1<\alpha_2+1$ the two models converge in the mathematical limit
$t\to\infty$, the speed of convergence is very different for cases (ii) and
(iii). Namely, when $\alpha_2<\alpha_1$, the deviation between MMFBM and MFBM
is dominated by the slower power-law $t^{-(1+\alpha_2-\alpha_2)}$, whereas
for $\alpha_2>\alpha_1$ the error decays to zero with the faster power-law
$t^{-1}$.}
\label{fig-msd-error}
\end{figure}

(ii) When $\alpha_2<\alpha_1<\alpha_2+1$, the ratio  (\ref{error}) decays
to zero with power $t^{-(1+\alpha_2-\alpha_1)}$ (an example is shown in
Fig.~(\ref{fig-msd-error})) and the MSD of MMFBM starts to converge to MFBM
after a time scale $\tau^{1/(1+\alpha_2-\alpha_1)}$. This time scale can
become much longer than $\tau$ as $1/(1+\alpha_2-\alpha_1)>1$, where we note
that in our dimensionless units, the time scale $\tau\gg1$, as unity
represents the elementary diffusive step.

(iii) When $\alpha_2>\alpha_1$, the ratio (\ref{error}) decays to zero
with power $t^{-1}$, and the MSD of MMFBM starts to converge to MSD after a
time scale which is equivalent to $\tau$.

\begin{figure}
\centering
(a)\includegraphics[width=0.47\linewidth]{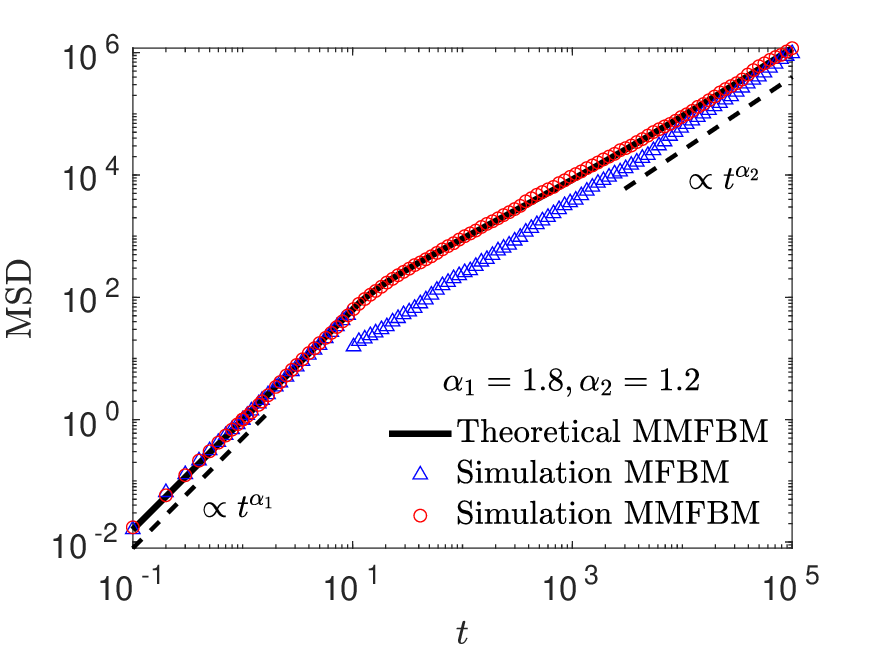}
(b)\includegraphics[width=0.47\linewidth]{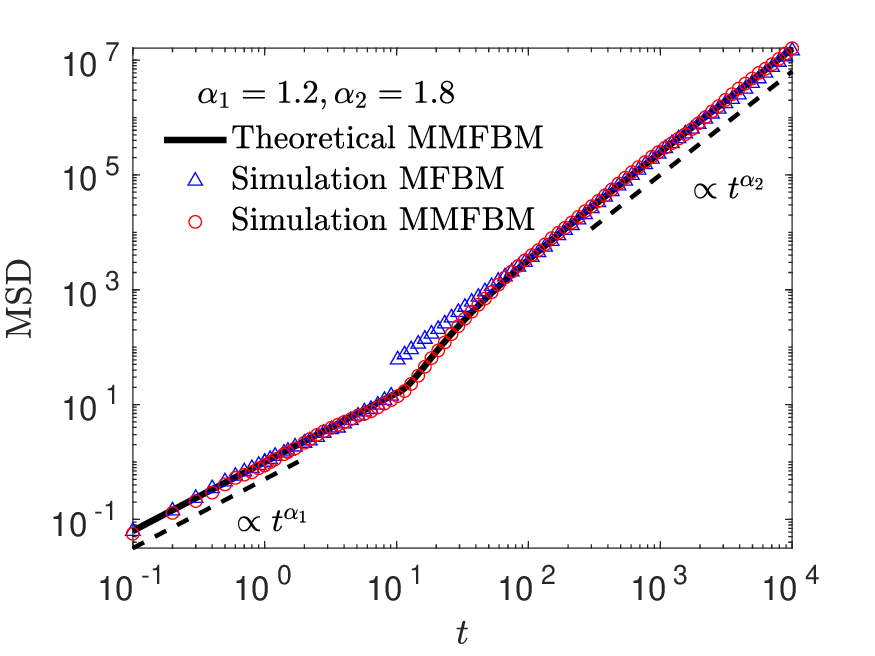}
\caption{MSDs for MMFBM and MFBM with switching time $\tau=10$. The theoretical
MSDs, Eq.~(5) in the main text, are represented by the full lines. In the left
panel, the MSD of MMFBM starts to converge to MFBM when $t\ge\tau^{1/(1+\alpha_2
-\alpha_1)}\approx316$, much longer than the switching time $\tau=10$, while in
the right panel the convergence time is equivalent to the switching time $\tau
=10$.}
\label{conv}
\end{figure}

We provide simulations results for cases (ii) and (iii) to validate the
physical limiting time observed for different choices of the exponents. In
Fig.~\ref{fig-msd-error} the much slower convergence of case (ii) is distinct.
It is thus necessary to go to extremely long times to observe the pure
$\alpha_2$-scaling of MFBM. For practical, physical applications such scales
can rarely be reached, and it is thus relevant to consider the memory
contained in MMFBM. We also show the MSD-convergence for two examples in
Fig.~\ref{conv}. For the case (ii) in the left panel, several orders of
magnitude in time need to be measured to observe convergence of the MMFBM
result to that of MFBM.

\section{Response function}

\begin{figure}
\includegraphics[width=0.47\linewidth]{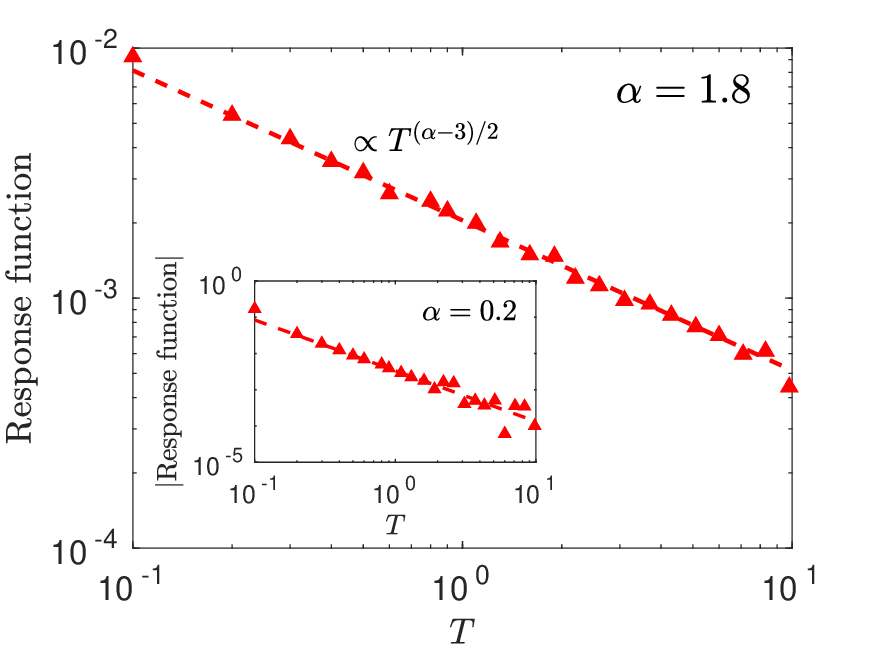}
\caption{Response function for MMFBM with switching time $\tau=1$. The dashed
curves represent the exact response function, Eq.~(2) in the main text. Unlike
the response function of MFBM, which is always zero ($T>0$), the response
function of  MMFBM decays with the power law $T^{(\alpha-3)/2}$.}
\label{resp}
\end{figure}

In Fig.~\ref{resp} we show an example for the response function.

\begin{thebibliography}{99}
\bibitem{REM} We here use a dimensionless notation. To restore physical units
the associated prefactor $\alpha(s)$ in Eq.~(\ref{mmfbm}) can be replaced by
the form $[K_{\alpha(s)}]=\alpha(s)\ell_0^2/t_0^{\alpha(s)}$. Dimensionality
can then be removed by setting the length scale $\ell_0$ and the time scale
$t_0$ to unity. The specific choice of the dimensionless prefactor $\alpha(s)$
in this expression for $K_{\alpha(s)}$ allows a compact notation and to
illustrate the most essential features of the MMFBM model.
\bibitem{REM1} We note that we choose the additional prefactor $\sqrt{\alpha(t)}$
in the definition of MFBM, compare \cite{ayache} in agreement with RL-FBM
\cite{mandelbrot}. This choice does not change the fundamental behavior.
\end{thebibliography}
\end{document}